\documentclass[twoside]{dis09}
\usepackage[latin1]{inputenc}
\usepackage[dvips]{graphicx,epsfig,color}
\usepackage{wrapfig,rotating}
\usepackage{amssymb,amsmath,array}

\begin{document}
\title{Status of MRST/MSTW PDF sets}

\author{\underline{R.S. Thorne}$^1$, A.D. Martin$^2$, W.J. Stirling$^3$ and G. Watt$^2$
%
\vspace{.3cm}\\
%
1-  Department of Physics and Astronomy, University College London, WC1E 6BT, UK \\
2-  Institute for Particle Physics Phenomenology, University of Durham, DH1 3LE, UK\\
3-  Cavendish Laboratory, University of Cambridge, CB3 0HE, UK \\
}

\maketitle

\begin{abstract}
  We outline the historical development of MRST/MSTW parton distribution
  functions (PDFs), and clarify how they should be regarded when compared
  to the most up-to-date 2008 MSTW sets, noting which sets are now obsolete
  and the reasons why.
\end{abstract}

\section{Status of PDFs}

Whenever a global fitting group produces a new set of PDFs it is
always appropriate to 
consider how these compare to previous sets, and indeed, to ask what is now the 
status of those previous sets, which may have already been used in some 
ongoing piece of analysis.  It is presumably always the case that the most 
recent set is the recommendation of the group, unless it corresponds to some 
atypical set provided for some specific purpose.  It is 
rarely the case that the previous set should immediately become regarded as 
obsolete, although there are some exceptions, but it is certainly true that 
older sets become more and more out-of-date, and at some point this fact 
alone should limit their use. 
The question of which PDF sets are still valid was discussed at the DIS 
2009 meeting and, for the case of MRST/MSTW, was addressed to some extent
in~\cite{url}.  Here we outline in more detail 
how the MRST/MSTW PDF sets have developed over 
the past decade, and give our views on the validity of each in the light of
the most up-to-date {\bf MSTW2008} sets~\cite{Martin:2009iq}.

\subsection{NLO PDF sets}

The first MRST PDF set was published in~\cite{Martin:1998sq}. This marked a 
transition from the previous MRS sets in a variety of ways. Not only was 
there an addition to the collaboration, but also the implementation of a 
general-mass variable flavour number scheme (GM-VFNS) (consistent with 
$\overline{\rm MS}$ evolution), instead of the more common zero-mass 
variable flavour number scheme (ZM-VFNS), and the inclusion of a wide 
variety of new data sets. The HERA structure function data had reached 
the point of being truly constraining at small $x$, E866 data added much 
more information on the $\bar u-\bar d$ difference, Tevatron data were 
starting to be included in a significant manner and fixed-target structure 
function data were still improving. In this article~\cite{Martin:1998sq}
NLO alone was considered, 
and as this is still the most widely used, we will first consider
the development of NLO sets. 


In principle, the {\bf MRST98} PDFs are mainly well-defined. The comparison to 
the structure function data, by far the major constraint, is made 
using a perfectly consistent GM-VFNS~\cite{Thorne:1997ga}. 
However, it was soon discovered that the 
evolution code contained a bug in the NLO $P_{gg}$ term (and a much less 
significant omission in the charm splitting functions), leading to mistakes of
order 1$\%$ in the gluon evolution which subsequently feeds through into, 
at worst, a similar error in quarks.  This is a shortcoming, though it 
should be borne in mind that the size of the error is
no larger than the uncertainty
quoted for the PDFs due to the accuracy of data even today.  A more serious 
shortcoming, in practice, is the fact that the data used have 
improved enormously
since this time, and the PDFs have changed significantly to account for this. 
In particular, in MRST98, prompt photon data were used to constrain
the high-$x$ 
gluon, and the early Tevatron jet data were only compared to qualitatively. 
Moreover, HERA structure function data quickly improved. 

In the {\bf MRST99}~\cite{Martin:1999ww} set the bug in the evolution code was 
corrected. However, the extraction of the central PDF sets was otherwise
essentially unchanged. The uncertainty of the PDF sets was estimated very 
crudely by a variety of model variations in the fit. Both the central sets 
and the range of variations should now be considered to be out-of-date.

The {\bf MRST2001} set~\cite{Martin:2001es} was a significant step forward. 
This included a fit to very much improved small-$x$, low-$Q^2$, 
data from HERA, 
indeed much the same as is still used.  Furthermore, the final Tevatron Run I
data on inclusive jet production were used, albeit via an indirect means of 
fitting, using pseudodata for the gluon with the correct $\chi^2$ calculated 
{\it a posteriori} (using a $K$-factor for the NLO cross section). 
Prompt photon data 
were dropped.  Data for $Q^2 \geq 2$ GeV$^2$ were fitted, and the best fit
required the addition of a negative $-A(1-x)^{\eta}x^{-\delta}$ contribution
to the input 
parameterisation for the gluon.  The MRST2001 PDFs are therefore fit to 
the majority of the data which provide the most important fundamental 
constraints for today's sets.  Moreover, they use a well-defined, though 
different, definition of a GM-VFNS.  Hence, although they are certainly 
far from as up-to-date as those of the MSTW2008 set, it is rather 
strong to view them as
obsolete. In general they are not too far from the MSTW central values. 
One major weakness is that the strange distribution is assumed to be a fixed 
fraction of the light sea rather than fit directly, and is a little high. 

The {\bf MRST2002} set~\cite{Martin:2002aw} was the first to be generated with 
quantitative uncertainties. The central set was fit to similar data 
as MRST2001 and is almost identical. 
The uncertainties were generated using the Hessian method 
developed in~\cite{Pumplin:2001ct}, with uncertainties being calculated using 
15 sets of orthogonal eigenvectors determined using a tolerance of an 
increase in global $\chi^2$ of 50 for an estimated $90\%$ confidence-level
uncertainty.  As with 
the central set there have been various changes in the meantime. The 
tolerance is now determined in a more sophisticated manner, but gives a 
rather similar result. There are more, and a better choice of, free parameters 
determining the eigenvectors (mainly for the strange and antistrange quarks, 
though there were some problems with the gluon and down valence 
distributions). Also, the uncertainty due to the data set normalisations was 
not considered. Overall, the uncertainty in the 2001 set was a little smaller 
than we would now obtain from fitting the same data, but by no more than 
a factor of two at the most, and usually rather less than this. Hence, as 
with the central 2001 set, we would suggest that the PDFs are qualitatively 
still acceptable, but one should be wary if requiring results with real 
precision. 

The {\bf MRST2003} set~\cite{Martin:2003sk} was obtained by increasing the 
data cuts at small $x$, $Q^2$ and $W^2$ until the quality of the fit stopped 
improving. The PDFs were therefore not constrained by data below $x=0.005$ and 
$Q^2=10$ GeV$^2$, and represented an extreme example of the effect of small-$x$
resummation and higher-twist corrections.  It would be 
interesting to repeat this study, but as an extreme example the 2003 set 
still plays the same r\^ole it always has. 

The {\bf MRST2004} analysis~\cite{Martin:2004ir} 
included a few new data sets, particularly some 
from HERA at moderate $x$ and high $Q^2$. However, its main change compared to 
the 2001 analysis was in the parameterisation of the gluon distribution. In an 
attempt to obtain the best simultaneous fit to HERA structure function data 
and the Tevatron Run I jet data, or 
at least to help explain the large high-$x$ gluon required, the gluon was 
parameterised in the normal manner in the DIS scheme.  
It then obtained a large 
contribution at high $x$ from the quarks via the transformation to 
the $\overline{\rm MS}$ scheme.  This leads to a distinctly different shape of 
the high-$x$ gluon compared to other MRST/MSTW sets.  The Tevatron Run II
jet data do not require such a large high-$x$ gluon.  Moreover, the
use of \textsc{fastnlo}~\cite{Kluge:2006xs}
(based on \textsc{nlojet++}~\cite{Nagy:2001fj,Nagy:2003tz}) and inclusion of 
hadronisation corrections mean that even
a fit only with Run I jet data does not favour the 
special 2004 parameterisation.  The MRST2004 sets are thus perhaps best
viewed as a specialised modification of the 2002 set.  No uncertainties
were provided 
for the MRST 2004 PDFs, but it is assumed they would be
very similar to the 2002 uncertainties.  Some
specialised sets, i.e.~with QED corrections~\cite{Martin:2004dh}
and in fixed-flavour schemes~\cite{Martin:2006qz}, 
are based on MRST2004, so it should be 
remembered that the high-$x$ gluon is rather large in these.
Updates will be forthcoming.   

The {\bf MRST2006} NLO PDF set was not publicly distributed, 
but complements the MRST2006 NNLO set~\cite{Martin:2007bv} which 
is discussed below. The NLO set differed from MRST2004 only 
in the application of an updated version of the GM-VFNS heavy flavour 
treatment. Hence, the difference between the PDFs and those of the
2004 NLO set is a measure 
of theoretical uncertainty due to heavy flavour schemes~\cite{Thorne:2008xf},
and can be a couple of percent.

\subsection{NNLO PDF sets}

The development of the NNLO PDF sets is more complicated than at NLO.  The
extent to which NNLO corrections have 
been calculated has increased significantly 
during the period over which the PDFs have been presented. Hence, there are 
more serious grounds for excluding some previous NNLO sets than is the case
at NLO.  


The first study of NNLO PDFs was performed in~\cite{Martin:2000gq}. 
This used approximate NNLO splitting
functions~\cite{vanNeerven:1999ca,vanNeerven:2000uj} obtained from
known small-$x$ 
limits and fixed-moment calculations~\cite{Larin:1996wd}. The study 
was distinctly approximate, with anticipated theoretical improvements and 
much improved HERA data imminent, so the PDFs were not made 
available. However, this was in fact the first study to 
include the second term in the gluon parameterisation, allowing it to go 
negative at very small $x$, since, as confirmed in all  
later MRST/MSTW studies, there is more requirement for the gluon to do this at 
NNLO than at NLO.     

The first MRST NNLO set to be distributed was
{\bf MRST2001}~\cite{Martin:2002dr}.  The {\bf MRST2002} NNLO
set~\cite{Martin:2002aw} is very similar.
This used improved approximations for NNLO 
splitting functions~\cite{vanNeerven:2000wp} made possible by the 
calculation of additional moments~\cite{Retey:2000nq}. It was fitted to
the same dataset as the MRST2001 NLO PDFs, and hence is acceptably complete. 
However, due to the approximate nature of the splitting functions the NNLO
analysis in~\cite{Martin:2000gq} and in~\cite{Martin:2002dr} used a
(fully explained) simplification of heavy flavour thresholds, 
which ignored discontinuities in both $\alpha_S$ and PDFs. The latter,
which awaited an updated definition of the GM-VFNS, turned out to be 
significant. The preliminary nature of this analysis was reflected in the 
fact that no error set was produced.  These NNLO PDFs are now obsolete. 

In the {\bf MRST2004} PDFs the full NNLO splitting
functions~\cite{Moch:2004pa,Vogt:2004mw} were used 
for the first time. Hence, no approximation was applied in this respect. 
However, since MRST2004 was essentially a specific modification of MRST2002
concentrating on the  high-$x$ gluon and Tevatron jets, no other significant 
improvements were made at NNLO, particularly in the case of heavy flavour. 
Hence, these sets are also now obsolete. 

For {\bf MRST2006} NNLO PDFs a full NNLO GM-VFNS was used for the first time
using the scheme of~\cite{Thorne:2006qt}, which combines elements
of~\cite{Thorne:1997ga} and~\cite{Tung:2001mv} and makes the explicit 
extension 
to NNLO. The NNLO Drell--Yan differential boson rapidity distributions 
also became
available~\cite{Anastasiou:2003yy,Anastasiou:2003ds}. This NNLO 
set is much more 
theoretically complete, and the 
production of uncertainty eigenvectors reflects this. The PDF grids were 
improved to deal with the PDF discontinuities and the opportunity was taken 
to extend them. It is a 
perfectly valid, though slightly out-of-date set, with the major unusual 
element being the scheme-transformation-inspired enhancement of the high-$x$ 
gluon which is the same as in the MRST2004 sets. This leads to a
peculiar feature in the uncertainty of the high-$x$ gluon
(see Fig.~17(b) of~\cite{Martin:2009iq}).

\subsection{LO PDF sets}

There have been fewer updates of LO sets, partially because of less demand 
for and interest in these, but also because some of the changes, e.g.~the
scheme transformation behind the MRST2004 sets, do not apply at LO.  Moreover,
the quality of the LO fit is not as good as it misses large higher-order
corrections to both coefficient and splitting functions.


The first LO MRST set was {\bf MRST98}~\cite{Martin:1998np}. It used the same 
approach as the MRST98 NLO set, but did not have the bug which occurred at 
NLO, and was therefore not updated in 1999. It is significantly out
of date due to 
the data used. 

The LO fit was updated in~\cite{Martin:2002dr}, producing {\bf MRST2001}. 
This is on a similar footing to the MRST2001 NLO sets, i.e.~somewhat out of 
date, but with no very significant features which would lead one to dismiss the 
PDFs as obsolete. However, no uncertainties were made available. 
As in the most up-to-date sets, a $K$-factor is needed to 
fit Drell--Yan data effectively and the fitted $\alpha_S$ values
are very large at LO.

\section{Conclusions}

\begin{wrapfigure}{r}{0.5\columnwidth}
\centerline{\includegraphics[width=0.5\columnwidth]{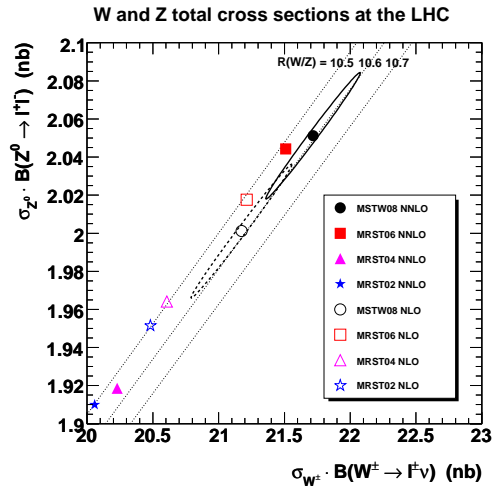}}
\caption{Predictions for $W$ and $Z$ cross sections at the LHC
for different PDF sets.}\label{Fig:WZ}
\end{wrapfigure}

We would, of course, recommend the use of {\bf MSTW2008} (LO, NLO, NNLO) PDFs
as default.  These 
contain the most up-to-date data sets, the broadest parameterisations of 
input PDFs, the most sophisticated treatment of heavy flavours, a heavy 
flavour scheme which is applied at NNLO, the fullest theoretical treatment 
in comparing data with calculations, a definition of the coupling which is 
identical to many other sets (and for the first time completely consistent 
at NNLO), and many other (in practice) minor improvements, e.g.~the 
tiny difference between quark and antiquark evolution at NNLO, leading to the 
perturbative generation of a $q-\bar q$ asymmetry for all flavours, is
applied for the first 
time.  However, few of the improvements at this, or previous stages of 
development, are sufficiently dramatic to require older versions of PDFs to
become obsolete.  The most up-to-date data leads to significant changes,
but all roughly in 
line with our previously quoted uncertainties. The most significant change in 
the 2008 PDFs from this source is to the high-$x$ gluon, but in practice it is
fairly similar to the high-$x$ gluon in the 2001 set.  The major improvement
in HERA data and 
Tevatron jet data around 2000 renders the pre-2000 PDFs of all groups obsolete
in our view.  As seen in~\cite{Tung:2006tb}, the error from using the
ZM-VFNS instead of a GM-VFNS can be large. 
Since MRST adopted a GM-VFNS from the outset, the changes at LO and 
NLO since then, from this source, are measures of theoretical uncertainty.  
At NNLO this aspect was treated correctly only from 2006 onwards, so 
MRST NNLO PDFs prior to this are obsolete.  The change in the coupling 
constant definition is of order a percent or so.  Our treatment of 
parameterisations 
and/or uncertainties has improved, but is qualitatively the same as in 
previous sets, except for the particular case of the strange distribution, so
if sensitivity to this is important then MSTW2008 is essential.  Our previous 
uncertainties are perhaps a little underestimated. 
A measure of the change in our PDFs is represented in~Fig.~\ref{Fig:WZ},
where one sees that 
the values of the $W$ and $Z$ cross sections at the LHC
are quite stable, except for the outlying pre-2006 NNLO sets, though the ratio 
varies a little more, at least in part due to the change in the strange quark 
contribution.

\begin{footnotesize}

\end{footnotesize}
\end{document}